\documentclass[aps,prx,reprint,amsmath,amssymb,longbibliography,floatfix,showpacs,showkeys]{revtex4-2}
\usepackage{graphicx}
\usepackage{placeins}
\usepackage{dcolumn}% Align table columns on decimal point
\usepackage{bm}% bold math
\usepackage{booktabs}
\usepackage{xcolor}
\usepackage[dvipsnames]{xcolor}
\usepackage{placeins}

\usepackage{natbib}
\begin{document}
\preprint{APS/123-QED}

\title{Contrasting one- and two--particle excitations in cuprate strange metals}

\author{Falk\,R\"oder}
\affiliation{Leibniz Institute for Solid State and Materials Research  Dresden, Helmholtzstr. 20, D-01069 Dresden, Germany}

\author{Roman\,Schuster}
\affiliation{Leibniz Institute for Solid State and Materials Research  Dresden, Helmholtzstr. 20, D-01069 Dresden, Germany}

\author{Martin\,Knupfer}
\affiliation{Leibniz Institute for Solid State and Materials Research  Dresden, Helmholtzstr. 20, D-01069 Dresden, Germany}

\author{Axel\,Lubk}
\affiliation{Leibniz Institute for Solid State and Materials Research  Dresden, Helmholtzstr. 20, D-01069 Dresden, Germany}
\affiliation{Institute of Solid State and Materials Physics, Technische Universität Dresden, Haeckelstr. 3, 01069 Dresden, Germany}

\author{J\"org\,Fink} 
\email{J.Fink@ifw-dresden.de}
\affiliation{Leibniz Institute for Solid State and Materials Research  Dresden, Helmholtzstr. 20, D-01069 Dresden, Germany}
\affiliation {Institute of Solid State and Materials Physics, Technische Universität Dresden, Haeckelstr. 3, 01069 Dresden, Germany}

\date{\today}
\begin{abstract}
In unconventional superconductivity, the pairing mechanism remains one of the central unresolved problems in solid-state physics. Moreover, the normal state shows strong deviations from conventional Fermi-liquid behavior, most notably a scattering rate that is linear in energy and temperature. Here we study the momentum dependence of the energy and damping of collective excitations in optimally doped cuprates in the normal state. In particular, we analyze existing Electron Energy--Loss Spectroscopy (EELS) data of Bi$_2$Sr$_2$CaCu$_2$O$_8$ [Phys. Rev. B 44, 7155 (1991)] and new analogous optical plasmon data of Bi$_2$Sr$_2$Ca$_2$Cu$_3$O$_{10}$. Furthermore, we reanalyze similar Resonant Inelastic X-ray Scattering (RIXS) acoustic plasmon data [Phys. Rev. R 6, 043184 (2024)] of La$_{1.84}$Sr$_{0.16}$CuO$_4$ and La$_{1.84}$Ce$_{0.16}$CuO$_4$. We contrast these two--particle excitation results with one--particle excitations measured by Angle-Resolved Photoemission Spectroscopy (ARPES). Most of the plasmon behavior can be described well by combined self-energies of a hole and an electron. The dispersion of optical plasmons at high energy is only weakly determined by correlation effects, indicating resilient quasiparticles. The damping of the optical plasmons at low energy, however, is strongly influenced by the reduced scattering rates described in the marginal Fermi liquid model for ``strange-metal'' behavior. In addition, the same model describes the linear-in-energy damping of acoustic plasmons. Contrary to some theoretical predictions, we see no overdamped optical plasmons below the classical Lindhard critical wave vector.
\end{abstract}

\pacs{74.72.-h,71.38.-k,79.20.Uv,71.45.Gm,79.60.-i}

\maketitle

\section{Introduction}\label{sec1}

Understanding how strong electronic correlations govern charge dynamics remains one of the central challenges of condensed-matter physics and a major obstacle to identify the  mechanism of high-$T_{\rm c}$ superconductivity. In the normal state, cuprate superconductors exhibit pronounced deviations from conventional Fermi-liquid behavior, most notably a ``strange-metal'' phase characterized by a resistivity that is linear in temperature over a broad range and persists even beyond the Mott--Ioffe--Regel (MIR) limit. Within a Drude picture, this limit corresponds to a mean--free path comparable to the lattice spacing~\cite{mott1972,ioffe1960,cooper2009}. Such anomalous transport behavior signals the breakdown of the quasiparticle paradigm and points to enhanced low-energy excitations and eventually to an unconventional continuum of low-energy electronic excitations, as captured phenomenologically by the marginal Fermi-liquid theory~\cite{varma1989}.

Spectroscopic probes provide a direct route to investigating this continuum and its influence on collective charge excitations. Yet recent experiments using different techniques have reached apparently conflicting conclusions regarding the existence, robustness, and correlation-induced renormalization of plasmons in cuprates. Resolving these discrepancies is essential for establishing a coherent description of charge dynamics in strongly correlated metals.

Recent holographic theories predict that the low-energy continuum, characteristic of strange metals, should strongly overdamp plasmons~\cite{Romero-Bermudez2019}. Experimentally, Electron Energy--Loss Spectroscopy (EELS) and Resonant Inelastic X-ray Scattering (RIXS) are particularly well suited to test this prediction because both probe the dynamic structure factor, which is directly related to the imaginary part of the charge susceptibility and contains information on both coherent collective modes and incoherent continua~\cite{giuliani2008,abbamonte2025}.

Early transmission EELS (T-EELS) studies of doped cuprates~\cite{Tarrio1988,Nuecker1989,Wang1990,Romberg1990,Nuecker1991,Knupfer1994} reported a well-defined optical plasmon near $\sim$ 1 eV with a positive momentum dispersion. For wave vectors below the $q_{\rm crit}$ which is the onset of the Lindhard continuum, the dispersion in Bi$_2$Sr$_2$CaCu$_2$O$_8$ (Bi2212) was found to be well described by the mean-field Random Phase Approximation (RPA), suggesting only weak correlation effects at these energies.

In contrast, recent RIXS measurements on hole-doped La$_{2-x}$Sr$_x$Cu$_4$O$_4$ (LSCO) and electron-doped La$_{2-x}$Sr$_x$Cu$_4$O$_4$ (LCCO)~\cite{Nag2024} revealed a pronounced renormalization of the acoustic plasmon velocity, indicating strong many-body effects through a reduced effective Fermi velocity. Moreover, reflection EELS (R-EELS)~\cite{Husain2019,chen2024} and very recent T-EELS data~\cite{devries2026} reported an optical plasmon only at very small momentum, followed by a rapid crossover into a broad continuum. In these experiments, the crossover occurs at momenta far below the classical threshold $q_{\rm crit}$ expected for the merging of the plasmon into the particle--hole continuum~\cite{giuliani2008}.

The FWHM energy width $\ddot\Gamma$  of the two--particle collective excitations (plasmons) is discussed since half a century. In many theoretical papers, the damping is only described by a phenomenological parameter. Originally, it was pointed out~\cite{DuBois1969} that plasmons at small momentum should not be damped  because,  according to the Lindhard model~\cite{Lindhard1954} in the long wavelength range, there  is no continuum of intra--band transitions which could mediate a decay of the plasmons. Later, in Ref.~\cite{Paasch1970} it was proposed that the plasmon damping at long wavelength is caused by a decay into inter--band transitions due to backfolded bands. For the nearly--free electron metals, this concept was experimentally confirmed by early EELS experiments on alkali metals which showed that the damping is determined by the square of the pseudopotential of the ions which causes the backfolding of the bands~\cite{Gibbons1977,Felde1989}. 

The  damping relative to the plasmon energy $\omega_{\rm p}$, $R_{\rm p}=\ddot\Gamma_{\rm p}/\omega_{\rm p}$ in these simple metals is rather small: for Na $R_{\rm p}\approx 0.05$ ~\cite{Felde1989} and for Al $R_{\rm p}\approx 0.03$~\cite{Raether1980}. However, in the doped cuprates and other strange metals,  $R_{\rm p}$ is of the order of 0.6. Although this number remains below unity, which is the nominal criterion for a well defined quasiparticle like collective mode, this raises the fundamental question: why do plasmons in cuprates remain observable despite the strong incoherent background implied by the strange-metal phenomenology.

A useful framework for addressing this question is to distinguish between one-particle and two-particle spectroscopies. In one-particle probes such as Angle-Resolved Photoemission Spectroscopy (ARPES)~\cite{Damascelli2003} and Scanning Tunneling Spectroscopy (STS)~\cite{fischer2007}, an electron or hole is added to or removed from the system, and its propagation is governed by the complex self--energy $\Sigma$. The real part  $\Sigma^{'}$ induces a mass renormalization, whereas the imaginary part
$\Sigma^{''}$ determines the quasiparticle lifetime $\tau$ and the corresponding scattering rate $\dot\Gamma=1/\tau$~\cite{Mahan2000}. In conventional Fermi liquids, due to the Pauli principle, the phase-space restrictions lead to a scattering rate that vanishes at the Fermi level and grows quadratically with energy. In contrast, ARPES studies of cuprates reveal a nearly linear energy dependence over a wide range, together with a high-energy saturation and the emergence of ``waterfall''-like dispersions~\cite{kaminski2005,valla2007,Graf2007}. The interpretation of the saturation of the scattering rates ranges from a finite bandwidth~\cite{varma1989}, an anisotropic scattering rate coupled to the MIR limit~\cite{Hussey2006}, or a coupling to bosonic excitations~\cite{valla2007}.

In two-particle spectroscopy such as infrared conductivity (IR), Raman, EELS, and Resonant Inelastic X-ray Scattering (RIXS), the connection to the single-particle self--energy is less direct because the measured response involves  particle and hole excitations. Nevertheless, previous work has shown that optical scattering rates extracted from IR can be quantitatively related to ARPES self--energies under appropriate assumptions~\cite{kaminski2000,Allen2015a,Allen2015}. A similar relationship between ARPES and Raman results has been worked out  in Ref.~\cite{Prestel2010}. Establishing such a connection offers a powerful route to reconciling apparently contradictory spectroscopic observations.

In previous T-EELS studies of doped cuprates, the dispersion of the optical plasmons was compared with calculations within the mean-field Random Phase Approximation (RPA) in the long-wavelength limit~\cite{Nuecker1991,Grigoryan1999,Knupfer1994,Roth2020}. Similar work on ruthenates has been published in Ref.~\cite{Schultz2025}. In the present work, we develop a unified interpretation of collective charge dynamics in cuprates by systematically connecting one-particle and two-particle spectroscopy through a self--energy-based analysis. This approximation is sometimes called the two--quasiparticle lifetime model (2QPL)~\cite{Green1985b,Allen2015a,Allen2015}. There, the energy-dependent self--energy of the hole in the occupied states and that of the electron in the unoccupied states are taken into account.

In this contribution we present new EELS data on Bi2212 to demonstrate that above $\approx$ 0.4 eV reliable loss data can be separated from the quasi--elastic peak. We re-examine transmission EELS data for the optical plasmon in optimally doped Bi2212 and compare them with more recent measurements on the trilayer compound Bi2223. We further relate these results to recent RIXS studies of acoustic plasmons in optimally hole-doped La$_{2-x}$Sr$_x$CuO$_4$ (LSCO) and electron-doped La$_{2-x}$Ce$_x$CuO$_4$ (LCCO)~\cite{Nag2024}.

Our analysis shows that the contrasting behavior of the momentum dependence of optical and acoustic plasmons arises naturally from the strong energy dependence of the single--particle self--energy. At higher energy, where mass renormalization is reduced and quasiparticle coherence is partially restored, optical plasmons remain robust. At low energy, the scattering rates for optical plasmons are reduced. The acoustic plasmons are strongly renormalized and damped by linear-in-energy scattering rates consistent with the marginal Fermi liquid model~\cite{varma1989}. This picture resolves long-standing inconsistencies between different spectroscopic probes and provides new constraints on theoretical descriptions of collective excitations in correlated electron systems and the normal state of high-$T_{\rm c}$ superconductors.

\section{Mass enhancement and lifetime broadening in one- and two-particle spectroscopy}\label{sec2}

\subsection{One-particle spectroscopy (ARPES)}\label{subsec1}

\begin{figure*}[!htbp]
\includegraphics[width=1\textwidth]{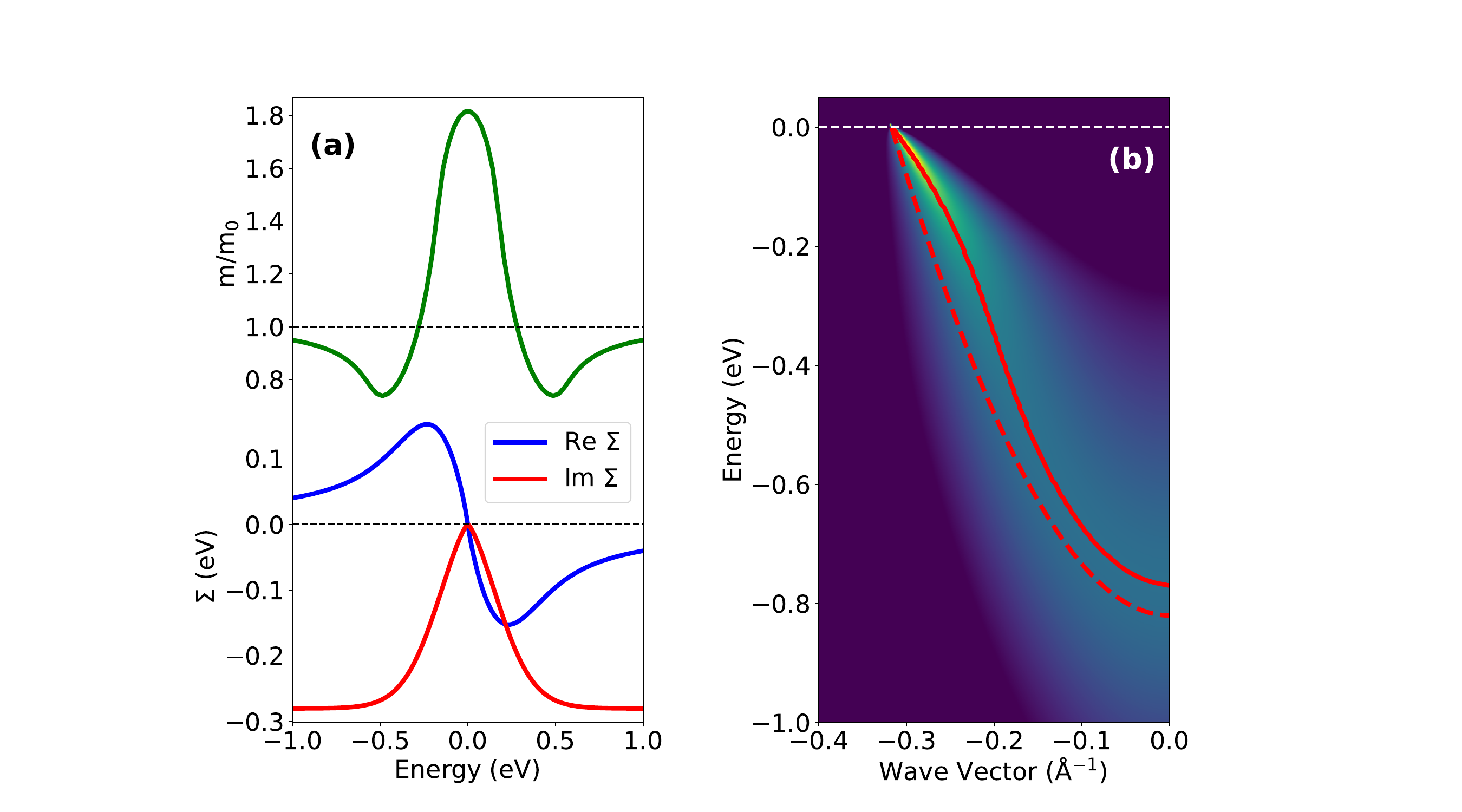}
\caption{ (color online) (a)  Schematic self--energy of a marginal Fermi liquid (after Ref.~\cite{kaminski2005}). 
The imaginary part of the self--energy $\Sigma''$ is presented by a red line.  The blue line shows the real part $\Sigma'$, obtained by Kramers--Kronig transformation. The green line indicates the effective mass $m^*/m_0$ calculated from Eq.~\ref{eq:eq4}.
(b) Spectral function calculated from the self--energy shown in (a) using Eq.~\ref{eq:eq1}. The dashed red line depicts the bare dispersion, and the solid red line the renormalized quasiparticle dispersion.}
\label{fig:fig1}
\end{figure*}

In one-particle spectroscopy such as ARPES and STS, the measured spectral function $A(\mathbf{k},\omega)$ is determined by the imaginary part of the retarded single-particle Green's function $G(\mathbf{k},\omega)$:

\begin{equation}
\label{eq:eq1}
A(\mathbf{k},\omega)= -\frac{1}{\pi}\Im\left[G(\mathbf{k},\omega)\right]= -\frac{1}{\pi}\Im\left[\frac{1}{\omega-\epsilon_{\mathbf{k}}-\Sigma(\omega)}\right].
\end{equation}

Here, $\epsilon_{\mathbf{k}}$ denotes the bare electronic dispersion 
and $\Sigma(\omega)$ the complex self--energy. Because local 
interactions dominate in highly correlated systems, $\Sigma$ is assumed
to be only weakly $k$ dependent. The real part, $\Sigma'(\omega)$, shifts the
occupied and unoccupied quasiparticle dispersion towards the Fermi energy and gives rise to an effective mass enhancement. The imaginary part, $\Sigma''(\omega)$, is related to the finite quasi--particle lifetime and thus is associated to the spectral broadening. The corresponding lifetime broadening (FWHM) of the single-particle excitation is given by
\begin{equation}
\label{eq:eq2}
\dot\Gamma = 2Z(\omega)\Sigma''(\omega).
\end{equation}
The quasiparticle weight $Z(\omega)$ is related  to the energy dependent inverse effective mass $m^*/m_0$.

There are numerous ARPES studies of the imaginary part of the self--energy in cuprates~\cite{kaminski2005,Kordyuk2005,valla2007,reber2019,smit2024}.
At low energy ($|\omega|\leq \Omega_{\rm sat}=$ 0.4 eV) the imaginary part of the self--energy is described by the relation
\begin{equation}
\label{eq:eq3}
\Sigma''(\omega)=a^{\rm ARPES}+b^{\rm ARPES}\omega.
\end{equation}
The parameter $a$ is related to elastic scattering. In the following  evaluations, this term is subtracted.  For Bi2212 and LSCO, we chose $b= $0.7 from Ref.~\cite{kaminski2005}. Using Eq.~\ref{eq:eq2} together with  the low-energy value $Z=0.5$, the energy dependent part of the scattering rate $\dot\Gamma(\omega<0.4\, \mathrm{eV})$ is equal to  $\Sigma''= 
b^{\rm ARPES}\omega$.
Along the nodal direction, above the characteristic energy $\Omega_{\rm sat}$ $\approx $ 0.4 eV, 
the scattering rate saturates, as expected within the marginal Fermi-liquid phenomenology.  For Bi2212 this yields with $Z$=1 (see below) above
$\Omega_{\rm sat}$ $\approx$ 0.4 eV the constant scattering rate $\dot\Gamma =  2Z\Sigma''$ = 0.56.

A similar saturation was also detected in optical spectroscopy~\cite{marel2003}. In Ref.~\cite{norman2006} this behavior was related to electrons interacting with a broad spectrum of bosons with a cutoff of the order of 300 meV. 

In cuprates, the local on--site Coulomb interaction dominates the correlation effects. That is the reason why the $k$-dependence of $\Sigma$ was detected to be weak~\cite{kaminski2005}. This is also part of  the phenomenological  Marginal Fermi Liquid theory~\cite{varma1989}. 

More recent ARPES data on LSCO revealed that at higher energies there is a strong $k$-dependence of the self--energy~\cite{chang2007,Chang2008}. Because the susceptibility of the charge excitations is mainly determined by excitations along the nodal direction, we neglect the angular dependence of the saturation energy.

A Kramers--Kronig transformation of $\Sigma''$ yields the corresponding real part $\Sigma'$. This function  reaches a maximum near $\Omega_{\rm sat}$, and decreases at higher energies. 
The energy dependent effective mass can be derived from $\Sigma'$ by the relation
\begin{equation}
\label{eq:eq4}
\frac{m^*(\omega)}{m}=\frac{1}{Z(\omega)}=1-\frac{\partial \Sigma'(\omega)}{\partial\omega}.
\end{equation}
It diverges at zero energy and reaches at high energy the value one. In Fig.~\ref{fig:fig1}~(a) we present
the functions $\Sigma'(\omega)$, $\Sigma''(\omega)$, and $m^*(\omega)/m_0$, which are derived from $\Sigma''(\omega)$ using ARPES results of Bi2212 from Ref.~\cite{kaminski2000}.

In Fig.~\ref{fig:fig1} (b) we present the calculated spectral function using the self--energy shown in Fig.~\ref{fig:fig1} (a).
 A strong low-energy renormalization is evident for $\omega<\Omega_{\rm sat}$, followed by a steep ``waterfall''-like dispersion, as observed experimentally~\cite{Graf2007,Meevasana2007}. At still higher energies, the renormalized band gradually approaches the bare dispersion, implying that the effective mass tends back towards its unrenormalized value one. This follows directly from Eq.~\ref{eq:eq4}.
 
The  waterfall structure, which is related to the saturation of $\Sigma''$ at $\Omega_{\rm sat}$, is is commonly interpreted as a consequence of the finite occupied bandwidth~\cite{Kotliar2006}. Alternative explanations invoke coupling to high-energy antiferromagnetic fluctuations extending up to $\sim 0.4$ eV~\cite{Letacon2011,Bok2010,Bok2016}. We note that qualitatively similar spectral functions can also arise from coupling to a dispersion-less Einstein phonon mode, as discussed in Ref.~\cite{engelsberg1963}. In that case, a comparable saturation would be expected once the relevant bosonic energy scale is exceeded.
Another explanation of a strong "kink" structure causing a "waterfall" dispersion was not based on a coupling to bosonic modes but  on a transition between the coherent band and the upper Hubbard band~\cite{byczuk2007}. We also note that  some ARPES studies attribute the waterfall feature primarily to matrix-element effects rather than many-body renormalization~\cite{inosov2007,rienks2014}.

\subsection{Two-particle spectroscopy (IR, EELS, and RIXS)}

In the simplest approximation, the complex optical conductivity $\sigma(\omega)$ can be described by a Drude--Lorentz model~\cite{Basov2005}:
\begin{equation}
\label{eq:eq5}
4\pi\sigma(\omega)=\frac{\omega_{\rm p}^{2}}{\ddot\Gamma-i\omega}+\sum_{L}\frac{\omega_{\rm pL}^2\omega}{i(\omega_{\rm L}^2-\omega^2)+\omega\ddot\Gamma_{\rm L}}.
\end{equation}
The first term describes intra--band charge dynamics in terms of the plasma frequency $\omega_{\rm p}$ and two-particle scattering rate $\ddot\Gamma$. The second term represents the sum of higher-energy excitations through Lorentz oscillators with strength $\omega_{\rm pL}$, resonance energy $\omega_{\rm L}$, and scattering rate $\ddot\Gamma_{\rm L}$. 
We emphasize, that different from one--particle spectroscopy,  where $\omega$ is related to the Fermi level, in  two--particle spectroscopy $\omega$ corresponds to the energy difference between hole and electron.

The optical conductivity is related to the complex dielectric function through~\cite{Wooten1972}
\begin{equation}
\label{eq:eq6}
\epsilon(\omega)=\epsilon_{\infty}+\frac{4\pi i}{\omega}\sigma(\omega)
\end{equation}
where $\epsilon_{\infty}$ accounts for high-energy screening processes. In EELS and RIXS, the relevant observable is the momentum- and energy-dependent loss function
\begin{equation}
\label{eq:eq7}
L(\mathbf{q},\omega)=\Im\left[-\frac{1}{\epsilon(\mathbf{q},\omega)}\right]\propto \chi''(\mathbf{q},\omega),
\end{equation}
which is proportional to the imaginary part of the charge susceptibility $\chi(\mathbf{q},\omega)$~\cite{Schnatterly1979,Fink1989,abbamonte2025}.
Within the Random Phase Approximation (RPA) the screened susceptibility $\chi(\mathbf{q},\omega)$ can be derived from the bare susceptibility
$\chi_0(\mathbf{q},\omega)$ by the relation
\begin{equation}
\label{eq:eq8}
\chi(\mathbf{q},\omega)= \frac{\chi_0(\mathbf{q},\omega)}{1-V(\mathbf{q})\chi_0(\mathbf{q},\omega)}
\end{equation}
with $V(\bf{q}) = $ 4$\pi e^2/q^2$.

In the Drude--Lorentz model, the loss function can be written as
\begin{equation}
\label{eq:eq9}
L(\omega)=\Im\left[\frac{1}{\epsilon_{\infty}-\dfrac{\omega_{\rm p}^{2}}{\omega^{2}+i\ddot\Gamma\omega}+\sum_{L}\dfrac{\omega_{\rm pL}^{2}}{\omega_{\rm L}^{2}-\omega^{2}-i\ddot\Gamma_{\rm L}\omega}}\right].
\end{equation}
At low frequencies, the Drude contribution reduces to
\begin{equation}
\label{eq:eq10}
L(\omega)\approx \frac{\ddot\Gamma\,\omega}{\omega_{\rm p}^{2}}.
\end{equation}
For a constant scattering rate, the loss function therefore rises linearly with energy. In contrast, if the scattering rate itself is linear in energy, as expected for a marginal Fermi liquid, the low-energy loss function acquires a quadratic dependence on $\omega$. For homogeneous nearly-free-electron metals, the momentum dependence of the plasmon energy can be described within the RPA. Below the onset of the Lindhard continuum, the plasmon dispersion is given by~\cite{Mahan2000}
\begin{equation}
\label{eq:eq11}
\begin{split}
\omega_{\rm p}(q)=\omega_{\rm p}(0)\sqrt{1+\frac{3\langle v_{\mathbf{q}}^{2}\rangle_{\rm FS}\mathbf{q}^{2}}{5\omega_{\rm p}(0)^{2}}}\\\overset{q\to 0}{\longrightarrow}\omega_{\rm p}(0)\left(1+\frac{3\langle v_{\mathbf{q}}^{2}\rangle_{\rm FS}\mathbf{q}^{2}}{10\omega_{\rm p}(0)^{2}}\right),
\end{split}
\end{equation}
where $\langle v_{\mathbf{q}}^{2}\rangle_{\rm FS}$ is the squared Fermi velocity projected on $\mathbf{q}$ and averaged over the Fermi surface. Because the plasmon dispersion is often small compared to the plasmon energy itself, the momentum dependence is frequently parametrized in quadratic form:
\begin{equation}
\label{eq:eq12}
\omega_{\rm p}(q)\approx \omega_{\rm p}(0)+A_{\mathbf{q}}{\mathbf{q}}^{2},
\end{equation}
with
\begin{equation}
\label{eq:eq13}
A_{\mathbf{q}}=\frac{3}{10\omega_{\rm p}(0)}\langle v_{\mathbf{q}}^{2}\rangle_{\rm FS}.
\end{equation}
In layered systems, the plasmon dispersion depends separately on the momentum components parallel ($q_{\parallel}$) and perpendicular ($q_{\perp}$) to the conducting planes.
The Coulomb potential in this case is given by~\cite{Fetter1974}
\begin{equation}
\label{eq:eq14}
V({\bf q})=V(q_{\parallel},q_{\perp})=2\pi e^2\frac{d}{q_{\parallel}} \frac{\sinh(q_{\parallel}d)}{\cosh(q_{\parallel}d)-\cos(q_{\perp}d)}.
\end{equation}
Using this potential leads then to a plasmon dispersion
\begin{equation}
\label{eq:eq15}
\omega_{\rm p}^{2}(q_{\parallel})=\frac{3\langle v_{\mathbf{q}}^{2}\rangle_{\rm FS}}{5}q_{\parallel}^{2}+\omega_{\rm p}^{2}(0)\frac{q_{\parallel}d}{2}\frac{\sinh(q_{\parallel}d)}{\cosh(q_{\parallel}d)-\cos(q_{\perp}d)},
\end{equation}
where $d$ is the inter--layer spacing. For $q_{\perp}=0$ and small $q_{\parallel}$, Eq.~\ref{eq:eq15} reduces to the conventional optical plasmon dispersion. For $q_{\perp}d=\pi$, charge oscillations in neighboring layers occur out of phase. In this case, the mode becomes gapless and is referred to as an acoustic plasmon:
\begin{equation}
\label{eq:eq16}
\omega_{\rm p}(q_{\parallel})=q_{\parallel}\sqrt{\frac{3}{5}\langle v_{\mathbf{q}}^{2}\rangle_{\rm FS}+\frac{\omega_{\rm p}^{2}(0)d^{2}}{4}}.
\end{equation}
\par
In correlated systems, the simple Drude picture must be generalized by replacing the constant scattering rate with an energy dependent complex memory function $M(\omega)$. Its real part describes an energy-dependent mass renormalization, while its imaginary part yields the optical scattering rate:
\begin{equation}
\label{eq:eq17}
M(\omega)=\omega\left(\frac{m^*(\omega)}{m}-1\right)+i\ddot\Gamma(\omega).
\end{equation}

In a series of papers~\cite{Green1985,Green1985b} the authors calculated the dynamical screening of the free--electron Lindhard susceptibility $\chi_0$ by the single particle self--energy. In their work for inelastic X-ray scattering studies this screening is determined by the sum of the self--energies of the hole and the electron. A similar theory for the optical conductivity was presented in Refs.~\cite{Allen2015a,Allen2015}. The one particle self--energy has also been used to renormalize the particle-hole correlation function of the Raman intensity \cite{Prestel2010}.We now pursue an analogous approach in our case. The marginal Fermi liquid self--energy $\Sigma(\omega)=\Sigma'(\omega)+i\,\Sigma''(\omega)$ is energy dependent only, with $\Sigma^{\prime(\prime\prime)}$ real and $\Sigma''(\omega)=\Sigma''(|\omega|)$.  For $\omega>0$, the renormalized particle-hole excitations, discarding particle-hole scattering, reside at energies

\begin{equation}
    \omega_{ph,\boldsymbol{kq}}\simeq \allowbreak\Delta E_{\boldsymbol{kq}}\allowbreak - i\frac{\ddot\Gamma_{\boldsymbol{k}}(\omega)}{2}
\end{equation}

with a real part
\begin{equation}
   \Delta E_{\boldsymbol{kq}}=\allowbreak \varepsilon_{\boldsymbol{k}+\boldsymbol{q}} - \allowbreak \varepsilon_{\boldsymbol{k}} - \left(\Sigma'(\varepsilon_{\boldsymbol{k}}+\omega)-\Sigma'(\varepsilon_{\boldsymbol{k}})\right) >0
\end{equation}

specifying the renormaliziation of the bare energy bands for ground ($\varepsilon_{\boldsymbol{k}}$) and excited state ($\varepsilon_{\boldsymbol{k}+\boldsymbol{q}}$) shifting them towards Fermi level. The imaginary part 

\begin{equation}
   \ddot\Gamma_{\boldsymbol{k}}(\omega)=\dot\Gamma(\varepsilon_{\boldsymbol{k}})+\dot\Gamma(\varepsilon_{\boldsymbol{k}}+\omega).
\end{equation}
describes the two particle lifetime broadening by approximately summing the single particle expressions for the occupied and unoccupied band.

\iffalse
for  and at $\omega^{<}_{ph,\boldsymbol{kq}}\simeq \allowbreak\Delta E_{\boldsymbol{kq}}\allowbreak - i[ \Sigma''(e_{\boldsymbol{k} + \boldsymbol{q}}-\allowbreak\omega) + \Sigma''(e_{\boldsymbol{k} + \boldsymbol{q}})\allowbreak]$ for $\Delta E_{\boldsymbol{kq}}<0$, respectively. The renormalized dispersion $e_{\mathbf{k}}=\epsilon_{\mathbf{k}}+\Sigma'(\epsilon_{\mathbf{k}})$ refers to the standard on-shell approximation. We assume inversion symmetry $e_{\boldsymbol{k}}=e_{-\boldsymbol{k}}$.
\fi

Using the preceding particle-hole energies we now approximate the dynamical density susceptibility by a Lindhard like function. To start, we discard the energy dependence of the scattering rate $\ddot\Gamma$, which is the energy broadening (FWHM) due to a finite life-time of the two-particle excited state and set
\begin{align}
\chi_{0}(\boldsymbol{q},\omega) & =n\sum_{\boldsymbol{k}}\frac{f(\varepsilon_{\boldsymbol{k}})-f(\varepsilon_{\boldsymbol{k}+\boldsymbol{q}})}{\omega-\Delta E_{\boldsymbol{kq}}+i\frac{\ddot\Gamma}{2}}\label{c1}\\
 & =n\sum_{\boldsymbol{k}}\frac{2f(\varepsilon_{\boldsymbol{k}})\Delta E_{\boldsymbol{kq}}}{(\omega+i\frac{\ddot\Gamma}{2})^{2}-\Delta E^{2}_{\boldsymbol{kq}}}\,.\label{c3}
\end{align}

The parameter $n$ reflects the electron density of the conduction band, which will be adjusted to match the experimental value for the plasmon energy in the optical limit and $f$ describes the Fermi-Dirac distribution function. For $\ddot\Gamma\rightarrow0^{+}$, Eq. (\ref{c3}) is referred to as Ehrenreich--Cohen formula \cite{Ehrenreich1959,Nuecker1991,Grigoryan1999}. Two comments are in order. First, density dynamics is subject to the continuity equation and particle number conservation. Therefore, it is noteworthy, that the susceptibility of Eqs. (\ref{c1},\ref{c3}) satisfies $\chi_{0}(\boldsymbol{0},\omega)=0,\,\forall\,\omega\neq0$ due to the form of their numerators, i.e., 
at least
the total particle number is conserved for finite $\ddot\Gamma$. Second, the spectrum $\chi''_{0}(\boldsymbol{q},\omega)$ of the susceptibility has to satisfy anti-symmetry, i.e., $\chi''_{0}(\boldsymbol{q},-\omega)=-\chi''_{0}(\boldsymbol{q},\omega)$.  For Eqs. (\ref{c1},\ref{c3}) this follows, using $\varepsilon_{\boldsymbol{k}}=\varepsilon_{-\boldsymbol{k}}$.

Next, we seek to extend the Lindhard function Eq. (\ref{c1}) by incorporating the energy dependencies of $\Sigma'(\omega)$ and $\Sigma''(\omega)$. There is no unique way to achieve this. First, regarding the real part $\Sigma'(\omega)$, we absorb this into the quasiparticle $Z$-factor, as usual. This renormalizes $\Delta E_{\boldsymbol{kq}}$, which we do not display explicitly. Moreover it replaces $\Sigma''(\omega)$ by $Z(\omega)\Sigma''(\omega)\equiv\dot\Gamma(\omega)/2$, as in Eq. (\ref{eq:eq2}). As one of the hallmark results of Fermi-Liquid theory, for conserved quantities, like the density in our case, the change of weight of the single particle Greens function induced by $Z$ does not have to be considered evaluating the dynamical susceptibility \cite{Leggett_1965}. Next, for the imaginary part $\Sigma''(\omega)$, one option is to considering $\omega>0$ and use $\omega^{>}_{ph,\boldsymbol{kq}}$ to write
\begin{equation}
\chi_{L+}(\boldsymbol{q},\omega)=n\sum_{\boldsymbol{k}}\frac{f(\varepsilon_{\boldsymbol{k}})-f(\varepsilon_{\boldsymbol{k}+\boldsymbol{q}})}{\omega-\Delta E_{\boldsymbol{kq}}+i\frac{\ddot\Gamma_{\boldsymbol{k}}(\omega)}{2}}\,,\label{c2}
\end{equation}
\begin{equation}
   \ddot\Gamma_{\boldsymbol{k}}(\omega) =  2(Z( \varepsilon_{\boldsymbol{k}} + \allowbreak\omega)\Sigma''(\varepsilon_{\boldsymbol{k}} + \allowbreak\omega)\allowbreak + Z( \varepsilon_{\boldsymbol{k}} )\Sigma''(\varepsilon_{\boldsymbol{k}})).
\label{eq:eq22}
\end {equation}
\iffalse
The imaginary parts $\ddot\Gamma_{\boldsymbol{k}}(\omega)$ and $\ddot\Gamma_{\boldsymbol{k}+\boldsymbol{q}}(-\omega)$ of the particle-hole self-energies for $\omega>0$ and $\omega<0$, respectively, are not identical in general. Therefore, it seems that for $\omega<0$ a different form of $\chi_{L0}(\boldsymbol{q},\omega)$ has to be used with $\omega^{<}_{ph,\boldsymbol{kq}}$ replacing $\omega^{>}_{ph,\boldsymbol{kq}}$.  However $\chi_{L0}(\boldsymbol{q},\omega)$ versus $\omega$, is a sum of poles. In the perturbative regime, $\Sigma''(\omega)/e_{\boldsymbol{k}}\ll1$, these are located at $\omega\approx e_{\boldsymbol{k}+\boldsymbol{q}}-e_{\boldsymbol{k}}$, which maps $\ddot\Gamma_{\boldsymbol{k}}(\omega)$ into $\ddot\Gamma_{\boldsymbol{k}+\boldsymbol{q}}(-\omega)$.  Therefore Eq. (\ref{c2}) can be used for all $\omega$. Moreover, by the same argument the spectrum is antisymmetric, In the perturbative regime, i.e., $\chi''_{L0}(\boldsymbol{q},-\omega)=-\chi''_{L0}(\boldsymbol{q},\omega)$.
\fi
By similar reasoning, one may also extend the Ehrenreich--Cohen formula to include the energy dependence of 
$\Sigma^{(\prime)\prime\prime}(\omega)$, e.g., for $\omega>0$
\begin{equation}
\chi_{EC+}(\boldsymbol{q},\omega)=n\sum_{\boldsymbol{k}}\frac{2f(\varepsilon_{\boldsymbol{k}})\Delta E_{\boldsymbol{kq}}}{(\omega+i\frac{\ddot\Gamma_{\boldsymbol{k}}(\omega)}{2})^{2}-\Delta E^{2}_{\boldsymbol{kq}}}\,.\label{eq:eq23}
\end{equation}
It is important to note, that Eqs. (\ref{c1},\ref{c3}) are analytically identical, however, Eqs. (\ref{c2},\ref{eq:eq23}) are \textit{not}. I.e., the inclusion of the energy dependence of $\Sigma^{(\prime)\prime\prime}(\omega)$ remains a phenomenological procedure to be tested against experiment.

Finally, we emphasize, that while the dressed susceptibilities in Eq.  (\ref{c2},\ref{eq:eq23}) have been set up to agree with the limit of non-interacting particles and to satisfy particle number conservation they do \textit{not} display proper hydrodynamic behavior at finite damping $\Sigma''(\omega)$. For that purpose, the susceptibility has to be determined by a conserving approximation \cite{Baym_Kadanoff_1961, Baym_1962}, i.e.,  by solving the Bethe-Salpeter equation for the density-vertex, or by using mode-coupling theory for the density and current \cite{Becker_Fulde_1987, Brenig_Becker_1988}.  Despite this deficiency, we show in the following sections that a consistent modeling of the plasmon spectra can be obtained.

In the present study, there are two energy regions. In the high energy range,  we record the optical plasmons using EELS. There, $\Sigma$, $Z$, and therefore $\ddot\Gamma$ is constant (see Fig.~\ref{fig:fig1}).  On the other hand, we have realized that the energy dependent low energy part of the self--energy has some influence on the damping of optical plasmons. In the low energy range, we discuss acoustic plasmon data. In this case, $\Sigma''$ is (see Eq.~\ref{eq:eq2}) linear in energy (relative to the Fermi level). Then using Eq.~\ref{eq:eq22} $\ddot\Gamma= b \omega$, i.e. for constant $Z$ it is linear in energy.

\subsection{Calculations}
 In previous investigations Eq.~\ref{eq:eq23} was used~\cite{Nuecker1991,Grigoryan1999} for the calculation of  $\chi_0(\mathbf{q},\omega)$. Here we use relation Eq.~\ref{c2} with reduced approximations.
For the underlying bare-band electronic structure, we employed a tight-binding parametrization of 
cuprates taken from 
Ref.~\cite{markiewicz2005}. 

 We used the following bare band parameters in units of eV: $\mu$=-0.36,  $t_0$=0.36, $t_1=$-0.1,  $t_2$=0.035 for Bi2212 and $\mu$=-0.24,  $t_0$=0.43, $t_1=$-0.04,  $t_2$=0.03 for LSCO~\cite{markiewicz2005}. For the in-plane isotropic  particle--hole symmetric self--energy, we used for both compounds the data presented in Fig.~\ref{fig:fig1} derived from ARPES data of Ref.~\cite{kaminski2005}. This is based on the fact that the self--energy of optimally doped LSCO is very close to that of Bi2212~\cite{Chang2008}.

The energy dependence of the imaginary part of the self--energy,  was set to $\Sigma''= b\omega$ with $b$=0.7 for $\omega \lesssim $ 0.4 eV and constant above this energy. $\Sigma'$ was derived from a Kramers-Kronig transformation (see Sec.~\ref{sec2}~A). The corresponding scattering rate of a single hole is given by $\dot\Gamma(\omega) = 2Z(\omega)\Sigma''(\omega)$.

For the calculations related to the optical plasmon we used in Eq.~\ref{eq:eq9} the standard 3D Coulomb potential. For the analogous calculation of the acoustic plasmon, we employed the Coulomb potential described in Eq.~\ref{eq:eq14}. Moreover, in the calculation for Bi2212 we approximated the bi-layer compound by a one-layer system.

In the two-particle spectroscopy (IR, IXS, RIXS, and EELS)  the life time of the excited state  is determined by the relaxation of the hole \textit{and} the electron towards the Fermi level. Using the bare particle band structure and the ARPES self--energy  we have calculated  the over the Brillouin zone averaged imaginary part of the unscreened susceptibility $\chi''_0$ depending on the excitation energy and the momentum using Eq.~\ref{c2}. Using Eq.~\ref{eq:eq23} we have received the same results. The calculations were performed for $\mathbf{q} \parallel $ to [110] and [100]. Afterwards we use Eq.~\ref{eq:eq7} to calculate the loss function. The plasmon energy and the width is derived from a  fit of a Lorentzian to the calculated loss function.

\section{Samples and experimental methods}\label{sec3}

The preparation and characterization of the flux-grown Bi2212 single crystals with $T_{\rm c} = $ 84 K are described in Ref.~\cite{zhao1988}.
Optimally doped single crystals of Bi2223 with $T_{\rm c} = $ 110 K were grown by the floating-zone method. Details of the preparation and characterization are given in Refs.~\cite{Giannini2004,Levalois2016}.
The properties of LSCO single crystals with a hole doping of $p =$ 16\% and $T_{\rm c} = $ 38 K are described in Refs.~\cite{Nag2020,Nag2024}. High-quality LCCO films with an electron doping of $p =$ 16\% and $T_{\rm c} = $ 7.9 K were grown on SrTiO$_3$ substrates via pulsed-laser deposition~\cite{Nag2024}. The RIXS measurements are described in Refs.~\cite{Nag2020,Nag2024}.

The EELS spectra were recorded at room temperature using a dedicated high-resolution electron energy-loss spectrometer operated at a primary electron energy of 170~keV~\cite{Fink1989,Roth2014}. The earlier measurements were performed in Karlsruhe with an energy resolution of $\Delta E_{\rm FWHM}=130$~meV and a momentum resolution of $\Delta q_{\rm FWHM}=$ 0.05~$\text{\AA}^{-1}$. In the Dresden setup, the energy and momentum resolution were improved to $\Delta E_{\rm FWHM}=80$~meV and $\Delta q_{\rm FWHM}=$ 0.04~$\text{\AA}^{-1}$, respectively. The energy resolution was derived from current scans of the intensity at zero momentum. The momentum resolution was obtained from scans across a Bragg reflection. We emphasize that the high-voltage ripple of the power supply ($\approx 20$~V) does not contribute to the measured energy resolution because monochromator and analyzer operate at the same high potential during the electron flight time. The sample itself is held at ground potential inside an ultra-high-vacuum chamber, which also minimizes surface contamination, particularly carbon deposition during measurement. Thin Bi2212 and Bi2223 samples were prepared by mechanical exfoliation from high-quality single crystals immediately prior to insertion into the vacuum system.

Experimental details of the RIXS study are described in Ref.~\cite{Nag2024}.

\section{Experimental results}\label{sec4}

\begin{figure} [!htbp]
\includegraphics[width=0.45\textwidth]{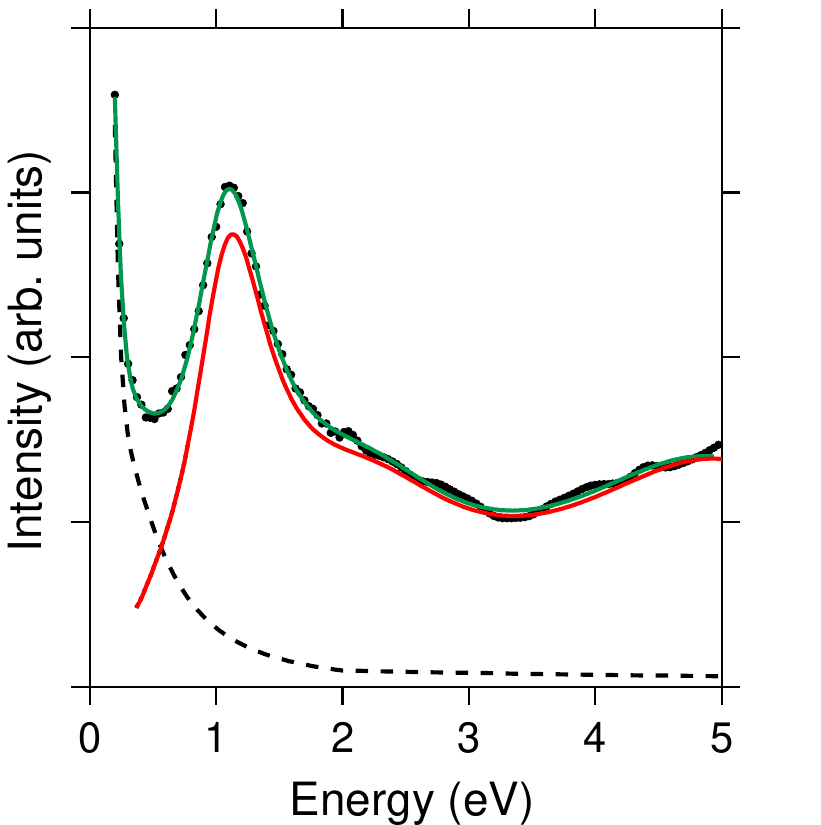}
\caption{(color online) Recent T-EELS spectra of Bi2212 for momentum transfer $\mathbf{q}\parallel[110] =$ 0.10~$\text{\AA}^{-1}$. Experimental data (black dots) are fitted by a Drude--Lorentz function plus a quasi-elastic contribution (green solid line). The quasi-elastic component is shown separately as a black dashed line. The corrected loss function, obtained after subtraction of the quasi-elastic peak is shown in red.}
\label{fig:fig2} 
\end{figure}

In Fig.~\ref{fig:fig2}, we present recent transmission EELS (T-EELS) data for optimally doped Bi2212. We remeasured these spectra to demonstrate that the loss function can be analyzed down to an energy of $\approx$ 0.5 eV, although at higher energies there is still contamination from the tails of the quasi--elastic peak. The spectra were analyzed using a Drude--Lorentz model with an energy-independent scattering rate. The low-energy quasi-elastic contribution was modeled by the sum of a Gaussian and a Lorentzian peak. Here we note that, with our setup, the quasi--elastic peak could not be measured together with the plasmon modes because there was a risk that the multiplier would be destroyed by an excessively high current.
The measured quasi-elastic peak is substantially broader than the instrumental energy resolution, $\Delta E_{\rm{FWHM}} =$ 0.08 eV measured by the energy dependent current of the direct beam. The additional broadening is most likely caused by unresolved low-energy phonon excitations. For the fit of the EELS data, we use Eq.~\ref{eq:eq9} with one plasmon mode and two Lorentz oscillators centered near 2.2 and 5.0 eV. The used background dielectric constant $\epsilon_{\infty}=4.5$~\cite{marel2003} is reduced   to 4.0 due to contributions of the 2 Lorentz oscillators. The fit yields an unscreened plasma frequency $\omega_{\rm p}^{*}=2.9$ eV, corresponding to a screened plasmon energy of $\omega_{\rm p}=1.2$ eV.

\begin{figure*}[!htbp]
\includegraphics[width=1. \textwidth]{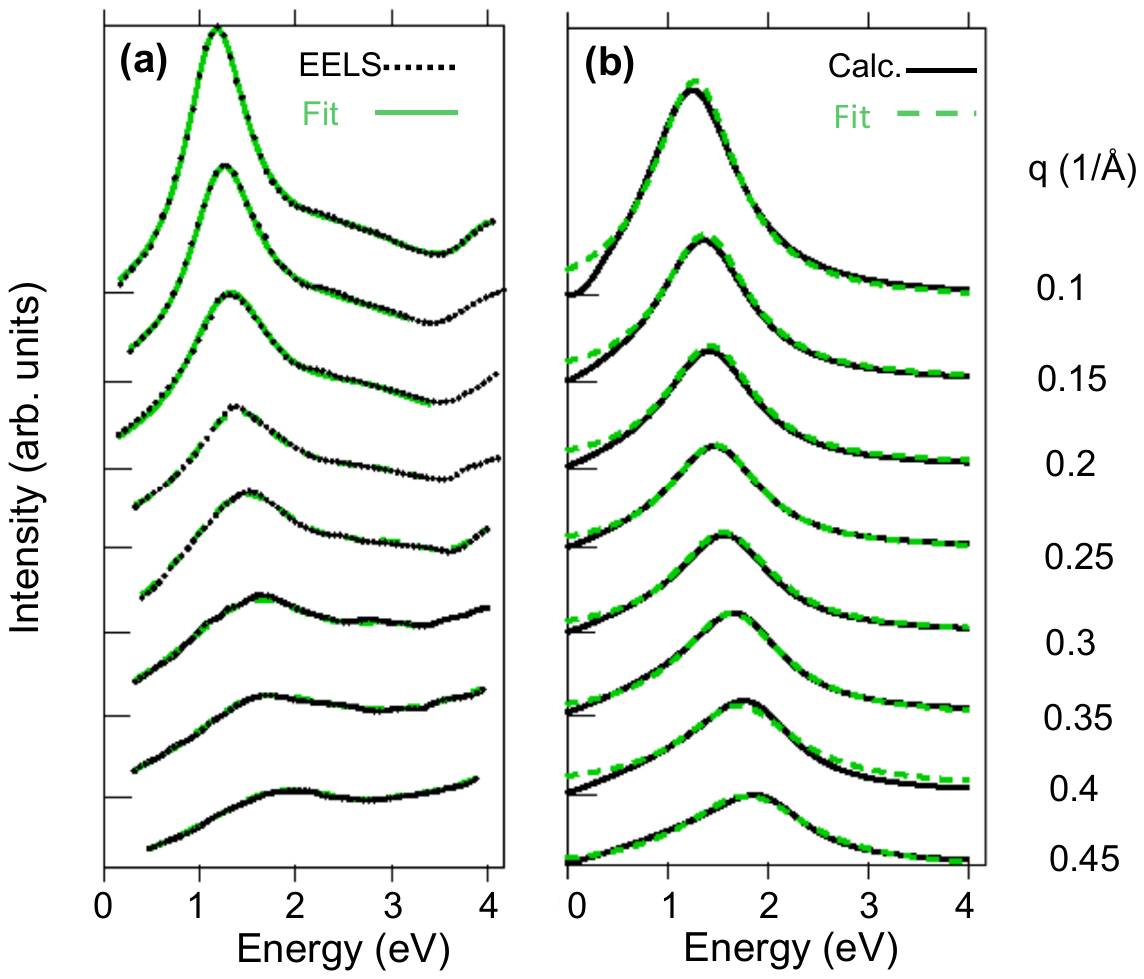}
\caption{(color online) Room temperature EELS results for Bi2212~\cite{Nuecker1989,Nuecker1991}. (a) Experimental loss spectra for momentum transfers between 0.10 and 0.45~$\text{\AA}^{-1}$ for $\mathbf{q}\parallel[110]$ (black lines), together with Drude--Lorentz fits (green dashed lines). (b) Fit of the plasmon component of (a) to a Lorentz function to derive the plasmon energy and the plasmon width.}
\label{fig:fig3}
\end{figure*}

\begin{figure*}[!htbp]
\includegraphics[width=1. \textwidth]{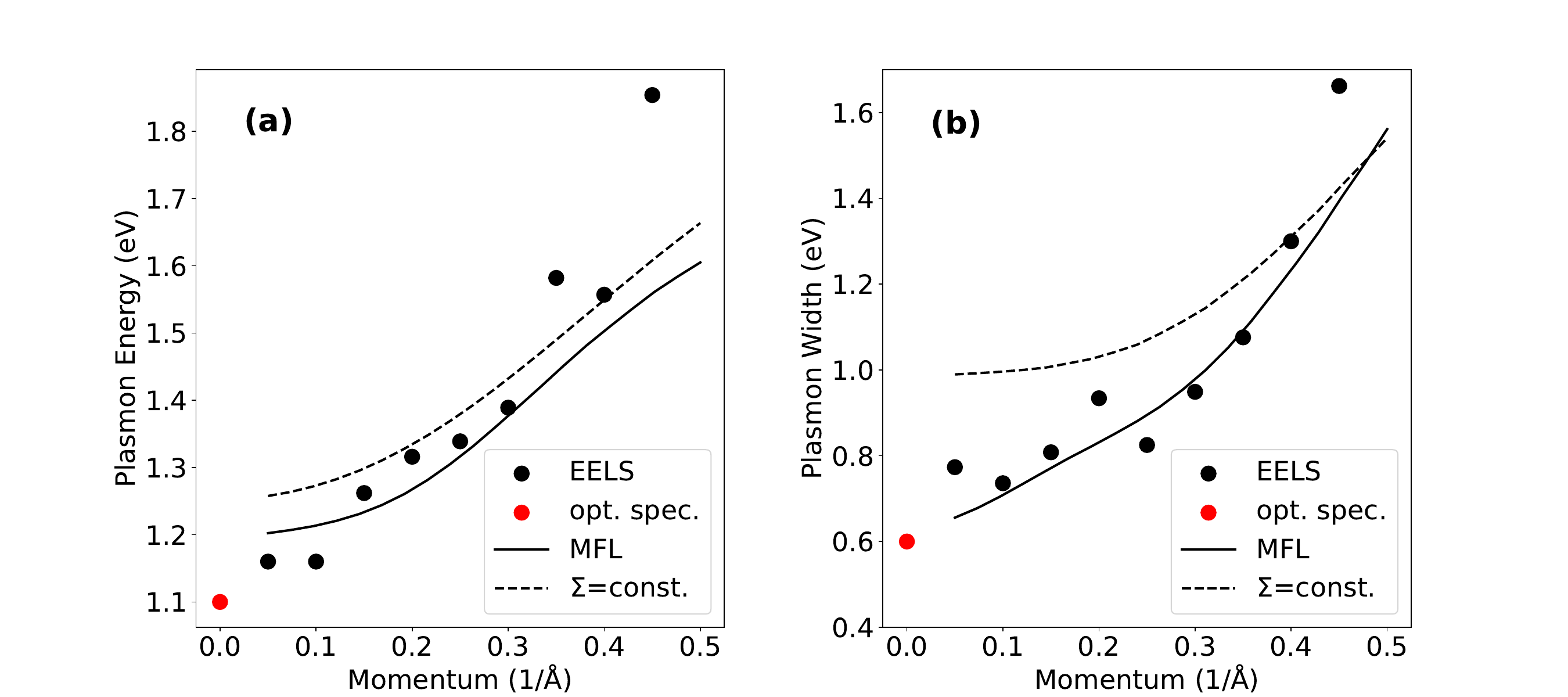}
\centering
\caption{ (color online) Room temperature EELS results of the plasmon dispersion (a) and the momentum dependence of the plasmon width (b) of Bi2212.  Black dots: experimental plasmon data extracted from the fits of the EELS data in Fig.~\ref{fig:fig3}~(a) and (b).  Solid black lines:  calculations taking a MFL self--energy  using   Eq.~\ref{c2}. Dashed black line: Calculations using a constant self--energy. Red dots: infrared data~\cite{Levalois2016} of the plasmon energy and the width are included. }
\label{fig:fig4}
\end{figure*}

In Fig.~\ref{fig:fig3}(a), we show the momentum-dependent loss spectra of optimally doped Bi2212 taken from Ref.~\cite{Nuecker1989} as a function of momentum transfer $\bf{q}$ parallel to [110], together with fits based on the Drude--Lorentz model. The quasi-elastic contribution was removed prior to the fitting procedure. In Fig.~\ref{fig:fig3}(b) we present the Drude part of the loss spectra from Fig.~\ref{fig:fig3}(a). The plasmon energy and width are derived by a Lorentzian fit to these data. The result of this analysis is shown in Fig.~\ref{fig:fig4}.
The experimental plasmon dispersion and width are close to the calculations using Eq.~\ref{c2}.

\begin{figure*}[!htbp]
\includegraphics[width=1.\textwidth]{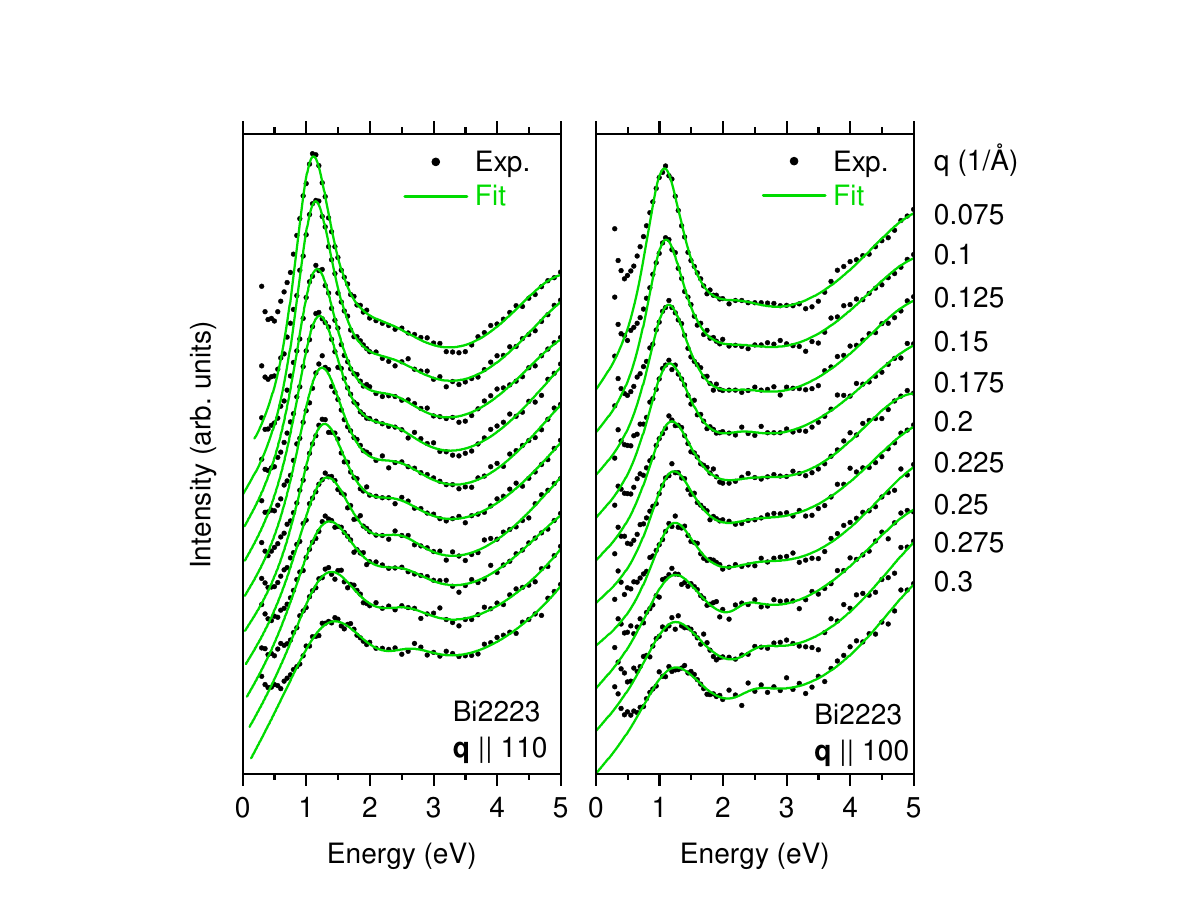}
\caption{(color online) EELS data from Bi2223 measured at room temperature. Black dots: experimental loss spectra for momentum transfers between 0.075 and 0.3~$ 1/\text{\AA}$. Green line: fit using a Drude--plasmon and two Lorentz oscillators. (a) Data for momentum $\mathbf{q}\parallel[110]$. (b) data for 
$\mathbf{q}\parallel[100]$.}
\label{fig:fig5}
\end{figure*}

\begin{figure*}[!t]
\includegraphics[width=1.0\textwidth]{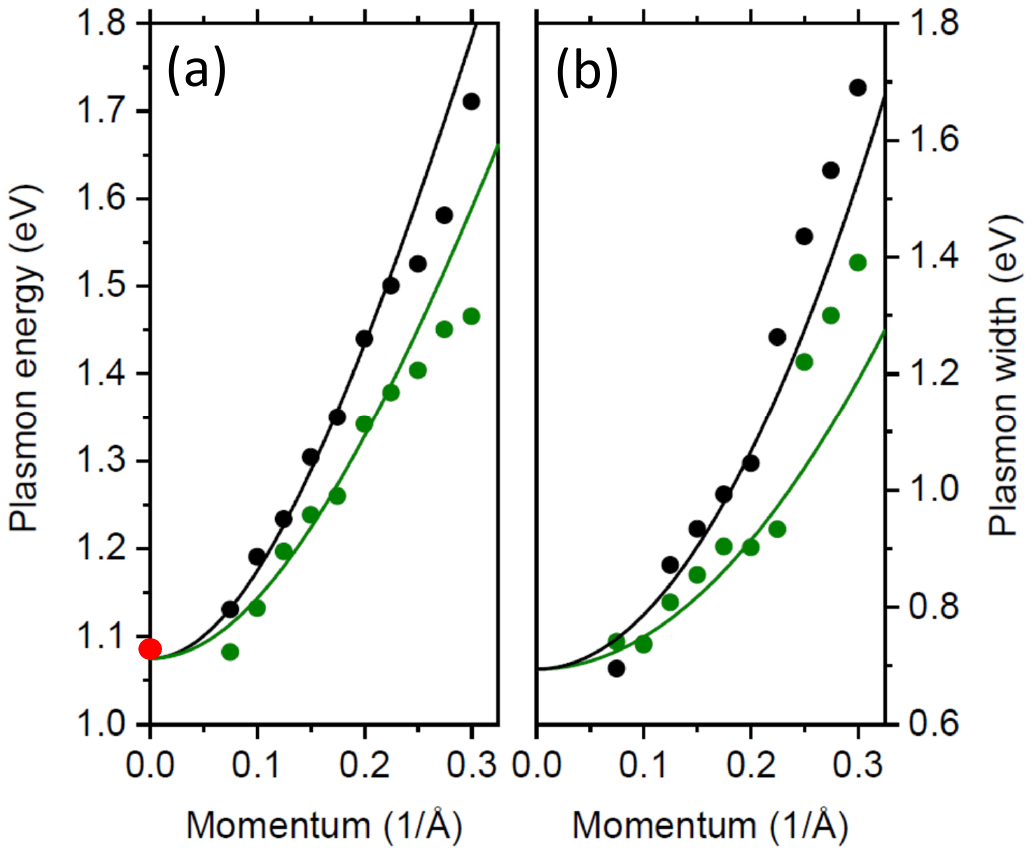}
\caption{ (color online) (a) Dispersion of the optical plasmon  from T--EELS data  of Bi2223. Black dots: data for $\mathbf{q}\parallel[110]$. Green dots: data for  $\mathbf{q}\parallel[100]$. (b)   Momentum dependence of the plasmon width. Lines: fit with a quadratic momentum dependence. Red dot: IR data from Ref.~ \cite{Levalois2016}. The same IR studies yield for the plasmon width $\approx$ 0.4 eV. Note the discontinuity of the plasmon width near $q= $0.25 $\text{\AA}^{-1}$}
\label{fig:fig6}
\end{figure*}

In Fig.~\ref{fig:fig5} we present recent T-EELS results from the tri--layer cuprate Bi2223 for two momentum directions. The plasmon energy and width are similar to those of the bi--layer compound Bi2212.
In Fig.~\ref{fig:fig6} we depict the plasmon energy and plasmon width as a function of momentum along $\mathbf{q}\parallel[110]$ and [100]. The data clearly show an anisotropy of the plasmon dispersion within the CuO$_2$ layer. For low momentum ($q\le 0.25~\text{\AA}^{-1}$), both the plasmon energy and width show a quadratic momentum dependence.

It is interesting to note that in the Bi2223 data a deviation from a quadratic momentum dependence is observed near $q = 0.25~\text{\AA}^{-1}$, well below the critical momentum $q_{\rm crit} \approx 0.45~\text{\AA}^{-1}$ derived for Bi2212. We also notice that the jump in the dispersion at $q = 0.25~\text{\AA}^{-1}$ is more pronounced along [100] than along [110].
\begin{figure*}
\includegraphics[width=1.0\textwidth]{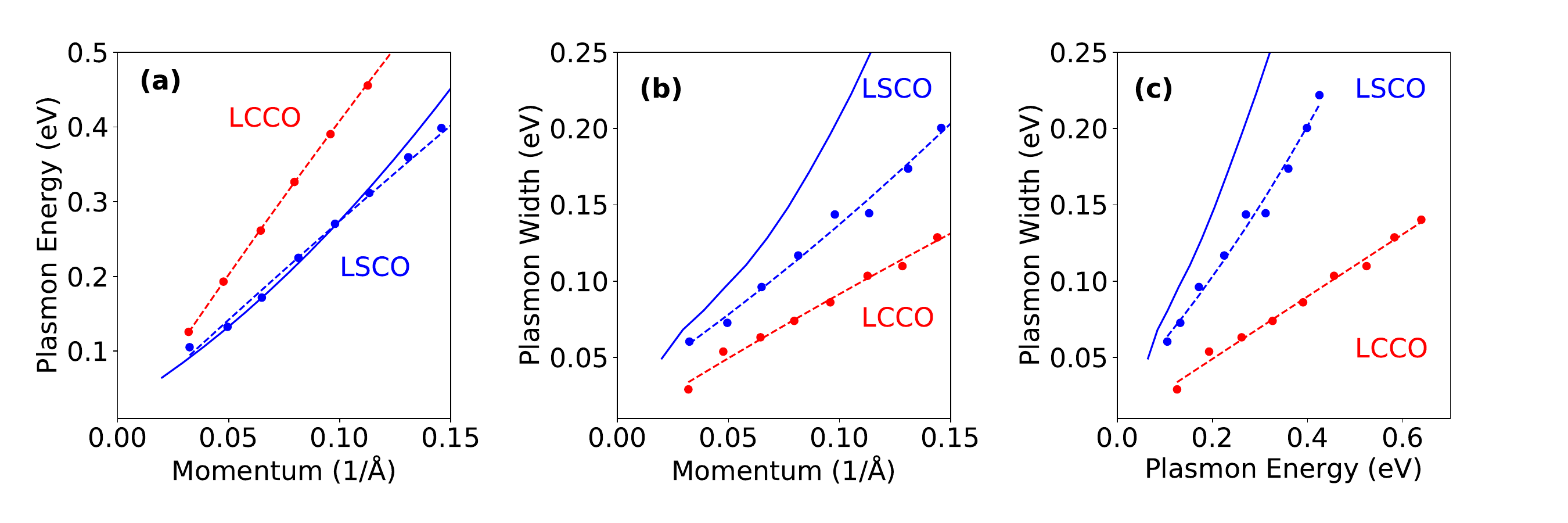}
\centering
\caption{ (color online) RIXS results of hole doped LSCO (blue dots) and electron doped LCCO (red dots)(Data taken from Ref.~\cite{Nag2024}). Dashed line: linear fit, solid line: calculations using Eq.~\ref{c2}  with energy dependent self--energies, quasiparticle residues, and effective mass. (a) Energy dependence of the acoustic plasmon as a function of momentum transfer $\mathbf{q}\parallel[100]$.  (b) Plasmon line-width as a function of momentum.
(c) Plasmon width as a function of plasmon energy.  }
\label{fig:fig7}
\end{figure*}

Next, we compare the high-energy optical plasmon data recorded by T-EELS with those of the low-energy acoustic plasmon measured by RIXS~\cite{Nag2024}. In Fig.~\ref{fig:fig7} the experimental data are compared with calculations using Eq.~\ref{c2}.
In Figure~\ref{fig:fig7}~(a) we show the acoustic plasmon dispersion for hole-doped LSCO and electron-doped LCCO.
The small finite offset at zero momentum likely reflects weak interlayer hopping. Likewise, the line-width extrapolated to zero energy remains below 0.01 eV, indicating only weak broadening from elastic scattering, inter-layer coupling, or a continuum.
Figure~\ref{fig:fig7}(b) shows the width of the acoustic plasmon as a function of momentum. Finally, in Fig.~\ref{fig:fig7}(c) we depict the derived plasmon width as a function of the plasmon energy. From these data we extract a \textit{linear} energy-dependence of the scattering rates for both compounds. For LSCO, we obtain $\Gamma(\omega) = b^{\rm RIXS} \omega$ with $b^{\rm RIXS}=$0.48, whereas for LCCO we find $b^{\rm RIXS}=$ 0.20.

\section{Discussion}\label{sec5}
We start  with the discussion of the bi-layer Bi2212 results which are shown in Fig.~\ref{fig:fig2} and \ref{fig:fig3} (a).  In  previous papers~\cite{Nuecker1991,Grigoryan1999}, the momentum dependence of the plasmon energy was analyzed on the basis of RPA calculations using a constant scattering rate. In this framework, the plasmon dispersion was compatible with an effective mass $m^*/m_0=$1, i.e., the plasmon dispersion is not renormalized by correlation effects. A calculation using an enhanced effective mass of two derived from low--energy ARPES experiments yields a considerable deviation from experiment. Similar results were also detected in EELS studies of the system Sr$_2$RuO$_4$~\cite{Schultz2025}, a compound which exhibits above 30 K a marginal Fermi liquid behavior.  This was explained in the following way: at high energy, scattering by boson excitation is essentially elastic. The electrons do not lose a significant fraction of their energy by low energy boson excitations. Thus, the bosons effectively create a static disordered background potential for the electrons. 

This view is supported by DFT+dynamical mean-field calculations on strongly correlated materials~\cite{deng2013} where resilient quasiparticles at high energy were predicted.
Moreover, an analysis of optical spectra of doped cuprates within the extended Drude-Lorentz model yields a reduction of the effective mass at higher energies~\cite{VanHeumen2007}. A  recent ARPES study of the quasiparticle spectral weight reports in another strange metal (Sr$_2$RuO$_4$) an increase of $Z$ with increasing temperature~\cite{Hunter2023}.

In this study,  we are taking it a step further. We take into account that in the one-particle spectroscopy ARPES, an energy dependence of the scattering rate has been detected. As outlined in Sec.~\ref{subsec1}, the imaginary part of the self--energy is energy dependent, which by the Kramers-Kronig transformation leads also to an energy dependence of the real part of $\Sigma$, the effective mass, and the quasiparticle residue $Z$.

Following the suggestion, that the single particle scattering rates are connected to the two-particle ones (see Sec.~\ref{sec2}). Using Eq.~\ref{c2} we have calculated the plasmon dispersion and the momentum dependence of the plasmon width.  The results of these calculations for the dispersion are not far from the experimental data (see Fig.~\ref{fig:fig4}~(a)). From this comparison, we conclude that the energy dependence of the self--energy has a small effect on the plasmon dispersion. Furthermore, the line shape of the loss function is close to a Lorentzian (see Fig.~\ref{fig:fig3}~(b)).

The discussion of momentum dependence of the  line width of the plasmon is quite different from that of the dispersion. First we assume a constant  scattering rate with an ARPES value  above $\Omega_{{\rm sat}}=$ 0.4 eV, which gives $\dot\Gamma=$2Z$\Sigma$= 0.56 eV ($Z=$1.0, $\Sigma=$0.28 eV). In the case of just adding the scattering rates of the hole and the excited electrons, we would expect a two--particle scattering rate $\ddot\Gamma$ and thus a FWHM plasmon width of 1.16 eV. This is in stark contrast to the EELS plasmon width of 0.65 eV, also detected in optical spectroscopy~\cite{Levalois2016}. From this we conclude that the plasmon width is related to the energy dependent imaginary part of the self--energy: the width is not only determined by the scattering rate near the plasmon energy, but it is also related to the marginal Fermi liquid behavior at low energy. This is understandable on the basis of Eq.~\ref{c2} describing the calculation of the bare susceptibility $\chi_0$. In the integral, at low energies, the scattering rate is more weighted than at high energies.

Using the energy dependent self--energy, also the momentum dependence of the plasmon width is reasonably reproduced (see Fig.~\ref{fig:fig4}~(b)). This gives us some confidence, that the used approximations described in Section~\ref{sec2} are not questionable.
It is astonishing that the loss function in the plasmon range, depicted in Figs.~\ref{fig:fig2} and \ref{fig:fig3}, can be fitted by a classical Drude--Lorentz model that assumes a constant $\ddot\Gamma$.
As outlined above, in particular the low-energy part of the loss function should be influenced by the energy dependence of the self--energy. With the present energy resolution and the tails of the quasi--elastic peak, this range is not accessible. On the other hand, this part was analyzed in a very early study of the optical properties of cuprates~\cite{bozovic1990,Bosovic1991} using reflectance and ellipsometry data. There, a non-Drude-like low-energy quadratic increase (with energy) of the loss function was discovered. This result was ascribed to excitations of low-energy acoustic plasmons. Our present analysis indicates that this observation was a very early hint of linear-in-energy scattering rates in optimally doped cuprates, supporting the phenomenological model of the marginal Fermi liquid~\cite{varma1989}. Using a linear energy dependence of $\ddot\Gamma$, Eq.~\ref{eq:eq10} yields a quadratic increase of the loss function.

The lifetime broadening of the plasmon ($\ddot\Gamma=$ 0.65 eV) is below the screened plasmon energy of 1.2 eV. This means that the damping is below the limit for the existence of a plasmon quasiparticle, which is set by the plasmon energy. We point out that our result is in contradiction with the theoretical prediction that in ``strange'' metals plasmons should be overdamped~\cite{Romero-Bermudez2019}.

On the other hand, at higher momentum near $q =$ 0.3 $\text{\AA}^{-1}$, the plasmon width is rather close to the plasmon energy (see Figs.~\ref{fig:fig3} and \ref{fig:fig4}). A calculation of the single--particle Lindhard continuum $\chi_0(q,\omega)$ indicates that near $q=$ 0.4 $\text{\AA}^{-1}$ the plasmon line is close to the upper end of the Lindhard continuum. This explains the rapid increase of the plasmon width near $q=$ 0.4 $\text{\AA}^{-1}$, a value that is close to the estimated $q_{\rm{crit}} = $ 0.45 $\text{\AA}^{-1}$. Moreover, the on-site Coulomb interaction $U$ may play an important role for short-wavelength plasmon excitations~\cite{prelovvsek1999}.

One would expect that the scattering rates measured in two-particle spectroscopy (IR and EELS) should be equivalent. We compare the optical value $\ddot\Gamma^{\rm {IR}}$(1 eV) $\approx$ 0.6 eV with the long-wavelength plasmon width of 0.65 eV. The difference is within the error bars.

The plasmon dispersion and the momentum-dependent plasmon width of the tri--layer compound are similar to those of the bi-layer system. On the other hand, the dispersion in Bi2223 is considerably reduced compared to that of Bi2212. This can be explained by NMR and ARPES results on tri--layer cuprates~\cite{trokiner1991,Kondo2017,ideta2025}. In these compounds there are inner and outer CuO$_2$ layers exhibiting different hole concentrations. For optimally doped Bi2223 the outer planes are slightly overdoped, while the inner layer is underdoped, which is related to stronger correlation effects. The latter may lead, even at higher energies, to a higher effective mass, thus causing a reduced plasmon dispersion.

The anisotropy of the dispersion is slightly reduced. The momentum dependence of the plasmon width for $\mathbf{q}\parallel[100]$ has a sudden step near $q$ = 0.25 $\text{\AA}^{-1}$. Possibly, the observation can be explained by the existence of charge-order fluctuations as derived from studies using resonant X-ray scattering, scanning tunneling spectroscopy, and time-dependent ARPES~\cite{Neto2014,Kunisada2020,smit2025}. We observe that the step near $q$ = 0.25 $\text{\AA}^{-1}$ is larger for the [100] direction when compared to that along [110]. This can be explained by the fact that stripes are normally detected along [100]/[010]. Moreover, $q$ = 0.25 $\text{\AA}^{-1}$ corresponds roughly to four times the lattice vector, which is close to the wavelength of stripe oscillations. On the other hand, the wave vector of the charge fluctuations detected in the ARPES studies is about a factor of two larger than that of the discontinuity realized in Fig.~\ref{fig:fig6}. Another explanation would be out-of-phase charge oscillations in the tri-layer compound~\cite{Nakata2025}.

A positive dispersion of the plasmon in Bi2212 along the momentum direction $\bf{q} \parallel$ [110] (nodal direction) was presented in Fig.~\ref{fig:fig3}(a).
The data definitely indicate that there is a plasmon up to a momentum $q$ = 0.45 $\text{\AA}^{-1}$. This value is close to the classical critical momentum $q_{\rm c} \approx \omega_{\rm p}/v_{\rm F} \approx$ 0.45 $\text{\AA}^{-1}$, at which the plasmon merges into the single particle-hole Lindhard continuum. The data were qualitatively reproduced in Ref.~\cite{Wang1990}. On the other hand, these results are at variance with R-EELS and T-EELS studies of other groups. There, transitions from a plasmon to a momentum-independent continuum were observed at $q$ = 0.26 $\text{\AA}^{-1}$~\cite{Mitrano2018,Husain2019}, at $q$ = 0.066 $\text{\AA}^{-1}$~\cite{chen2024}, and at $q$ = 0.15 $\text{\AA}^{-1}$~\cite{devries2026}. Other T-EELS studies~\cite{Terauchi1995,terauchi1999} reported a plasmon at $\sim$ 0.6 eV or just a continuum. The latter result is probably caused by a reduced momentum resolution, leading to a sampling of spectral weight of the entire BZ, i.e., both in the coherent range for $|\bf{q}|< q_{\rm{crit}}$ and in the range of the classical single-particle Lindhard continuum.

Another important result is the observation of an in-plane anisotropic plasmon dispersion. Analyzing the plasmon dispersion using Eqs.~\ref{eq:eq12} and \ref{eq:eq13} yields for the dispersion coefficient along the [110] direction $A_{\rm {[110]}}$ = 5.7 $\text{\AA}^{-2}$, while that along the [100] direction is reduced to $A_{\rm {[100]}}$ = 2.7 $\text{\AA}^{-2}$~\cite{Nuecker1991}. As described in Eqs.~\ref{eq:eq12} and \ref{eq:eq13}, in RPA the dispersion coefficient is proportional to $<v^2_{\rm {\bf{q}}}>$. We have calculated the averaged Fermi velocity for the two momentum directions (for details see Section~\ref{sec2}). The result is $<v^2_{\rm {[110]}}>$ = 7.2 (eV~\text{\AA})$^2$ and $<v^2_{\rm {[100]}}>$ = 3.7 (eV~\text{\AA})$^2$. The calculated ratio of the average squared Fermi velocities is then 1.9. It is close to the experimental ratio of the measured dispersion coefficients, 2.1.

In summary, our findings on the optical plasmon support the view that the plasmon damping in these strange metals is caused by the relaxation of electrons and holes toward the Fermi level, and not by decay into interband transitions, as revealed in nearly-free-electron metals~\cite{Paasch1970,Gibbons1977,Sturm1981,Felde1989}.

The anisotropic dispersion is related to the potential of the ions along $\bf{q}$, leading to a particular averaged Fermi velocity. It is not related to an incoherent electron liquid in which the dynamics of the electrons is determined by electron--electron interaction. This result can be rationalized by the fact that the plasmon energy is well above the energetically highest bosonic excitations, which are probably related to spin fluctuations ranging to 0.4 eV~\cite{norman2006}, or above the transition from a renormalized band structure to the upper Hubbard band~\cite{byczuk2007}. 

Next, we discuss the acoustic plasmon data, measured for  momentum parallel to the [100] direction. The experimental data are compared to calculations using  Eq.~\ref{c2} with an energy dependent ARPES self--energy and a Coulomb potential for a layer system (Eq.~\ref{eq:eq14}).  At low  momentum the calculation is rather well reproduced by a  Lorentzian in agreement with
the fact, that the convolution of two Lorentzians result again in a Lorentzian. At higher momentum the spectral weight is transferring into an asymmetric shape. At still higher momentum the difference between experiment and calculation increases. For LSCO and LCCO the RIXS results for the dispersion are presented in Fig.~\ref{fig:fig7}(a), together with the calculated dispersion. In Fig.~\ref{fig:fig7}(b) we depict the width as a function of the momentum. In Fig.~7(c) we show the width as a function of the plasmon energy for the two compounds together with the results of a calculation using Eq.\ref{c2}.

Calculations for the acoustic plasmons are only shown for the p-type-doped cuprate. The reason for this is that for the n-type-doped cuprates, there are no reliable self--energy data over a larger energy range. 
 
The acoustic plasmons are caused by anti-phase  collective excitations of valence charge density between adjacent layers separated by the distance $d$. From the RIXS results we estimated the importance of many-body effects relative to an independent particle approach~\cite{Nag2024}. In a weak--coupling mean--field RPA the acoustic plasmon dispersion as a function of $q_{\parallel}$ for a fixed $q_{\perp}= \pi/ d $ can be described by Eq.~\ref{eq:eq15}. The difference of the plasmon velocity $v_{\rm{p}}=\omega_{\rm{p}}/\rm{q}$ between LSCO and NCCO has been explained by the different averaged Fermi velocities which are smaller for the half-filled p-type doped compound compared to the larger $v_{\rm{F}}$ in the more than half-filled band structure of the n-type doped system. A more quantitative analysis yields that the plasmon velocity cannot be described by the bare-particle band structure but needs a mass renormalization $m^*/m_0 \approx$ 2. This value is close to the mass renormalization derived from ARPES. This means that the acoustic plasmon dispersion at low energy, different from the high-energy optical plasmon dispersion, is renormalized by many-body effects. A more refined analysis using  the strong-coupling t-J-V model in a large-N scheme, which describes correlation effects implicitly, supports the impact of electron correlations on the low-energy plasmon excitations~\cite{Nag2024}.
\par
The width of the acoustic plasmon has been explained, in analogy to simple metals, in terms of a decay into interband transitions caused by back-folding due to the finite pseudopotential of the ions~\cite{Nag2024}. By comparison with related low-energy scattering-rate data from IR spectroscopy and ARPES~\cite{kaminski2000}, in the present contribution we were led to a different interpretation.

Using Eq.~\ref{c2} for low energies and a linear energy dependence of $\dot\Gamma=b\omega$, we obtain for the two--particle lifetime broadening $\ddot\Gamma=b\omega$ as well. From ARPES we obtain for the slope $b$ a value of 0.7 (see Section~\ref{sec2}). The RIXS plasmon width vs. energy exhibits a slope of $b=0.5$ (see Fig.~\ref{fig:fig7}c). For LSCO the $b$ values from the two-particle spectroscopies IR and RIXS are, as expected, almost equivalent. The ARPES result $b=0.7$, derived from momentum-distribution curves in Ref.~\cite{kaminski2005}, is slightly larger.
On the other hand, ARPES studies based on energy-distribution curves yielded $b=0.4$. At present, it seems that within error bars the $b$ values from one- and two-particle spectroscopies are nearly the same. The small reduction of the experimental $b$ value compared to the calculated one may be explained on the basis of the assumed approximations (see Sect.~\ref{sec2}B), by vertex corrections and excitonic effects, or by the anisotropic saturation energy of $\Sigma''$ detected in ARPES data of LSCO~\cite{Chang2008}.

Calculations on the basis of Eq.~\ref{c2} yield a Lorentzian shape for low-energy loss spectra. The linewidth broadening is \textit{linear} in energy: $\ddot\Gamma=b\omega$ with $b=0.8$. Thus, the calculated scattering rates qualitatively agree with the experimental data (see Fig.~\ref{fig:fig7}(c)). At higher energies a shoulder appears in the calculated loss function, causing an increase of the width in the calculations. This is probably caused by a transition to the high-energy behavior of the scattering rate, by asymmetric spectral functions at higher energies, or by an artifact of the approximations used.

The $b$ factors of NCCO do not fit into this scheme. $b^{\rm IR}$ is almost a factor of two larger than that from RIXS, and the RIXS and ARPES values are almost equivalent.
The reduction of $b^{\rm RIXS}$ can possibly be explained by the different filling of the conduction band. In LSCO this band is nearly half filled, while there is a 70\% filling in NCCO~\cite{Nag2024}. In the two-particle spectroscopy RIXS, the scattering rate should be related to the joint density of occupied and unoccupied states. The density of states in the two-dimensional band structure of cuprates is nearly constant. Then the $n$-type-doped cuprates with the higher band filling should have a reduced scattering rate compared to the $p$-type-doped system with a half-filled band. Here we note that the reduced scattering rate, averaged over the hole Brillouin zone, may explain the reduced superconducting transition temperature $T_c$ of $n$-type-doped cuprates.

In this context, it is remarkable that the scattering rates derived from the linear-in-temperature resistivity are much larger in hole-doped cuprates than in electron-doped compounds~\cite{legros2019}.

\section{Conclusions}\label{sec6}
In the present study of collective plasmon excitations in optimally doped cuprates, we first demonstrated that, using an energy resolution of $\sim$ 0.1 eV, it is possible to extract meaningful results on plasmon excitations, although at low energies there is contamination from the quasi--elastic peak.

Second, we reevaluated early loss data on Bi2212 using the two--quasiparticle lifetime model. In Bi2212, the plasmon dispersion over the full range below the Lindhard continuum can be explained by an \textit{unrenormalized} electronic band structure, indicating resilient quasiparticles in the energy range of optical plasmons near 1 eV. On the other hand, the long--wavelength value of the plasmon width and its momentum dependence can only be understood on the basis of an energy-dependent, marginal-Fermi-liquid-like self--energy.

For Bi2223 we observe similar results in the low-momentum range. On the other hand, for larger momentum (but still below the critical wave vector of the Lindhard model) we detect deviations from the classical behavior. These may be related to out--of--phase oscillations in the tri--layer system or to charge-density fluctuations in the central underdoped CuO layer.

The finding of an unrenormalized dispersion for high-energy optical plasmons contrasts with the renormalized dispersion observed for acoustic plasmons at low energy. The widths of optical and acoustic plasmon excitations are explained within the quasiparticle lifetime model using an energy-dependent self--energy from ARPES experiments. This demonstrates that the damping of acoustic plasmons in these correlated ``strange'' metals is determined by the on-site Coulomb interaction on the Cu ions. This result differs from the damping of plasmons in simple metals, where decay into interband transitions is prevalent. In the momentum range below the classical Lindhard continuum, we do not observe any overdamped plasmons.

Finally, our analysis of the RIXS data on strange metals demonstrates that valuable information on scattering rates can be derived from RIXS or from future high-resolution EELS studies.

\section{Acknowledgements}
We thank Sergey Borisenko, Rudolf Hackl, Karsten Held, Dirk van der Marel, Thomas Roth, Matthias Vojta, and Krzysztof Wohlfeld for helpful discussions. We especially acknowledge theoretical support from Wolfram Brenig.
We thank Enrico Giannini for providing the Bi2223 crystals. Technical support by Marco Naumann and Roland H\"ubel is gratefully acknowledged. A.L. acknowledges financial support by the Deutsche Forschungsgemeinschaft (DFG) through CRC 1415 (Project No.~417590517). This project has also received funding from the European Union's Horizon Europe research and innovation programme under grant agreement No.~101094299. 

\section {DATA Availability}

All data supporting the findings are provided as figures in the article. Data files for all figures are available from the corresponding authors on
request.

%\bibliographystyle{apsrev4-2}
%\raggedright
\bibliography{sn-bibliography}

\end{document}